\documentclass[12pt]{article}

\usepackage{a4wide}
\usepackage{amsmath}
\usepackage{amsthm}

\usepackage{amsfonts}
\usepackage{amssymb}
\usepackage{bbm}
\usepackage{cite}
\usepackage{graphicx}

\newcommand{\N}{\mathbbm{N}}

\newcommand{\R}{\mathbbm{R}}

\renewcommand{\l}{\lambda}
\renewcommand{\L}{\Lambda}
\renewcommand{\b}{\beta}
\newcommand{\s}{\sigma}

\newcommand{\ie}{i.e.\ }
\newcommand{\e}{\mathrm{e}}

\newtheorem{theorem}{Theorem}[section]
\newtheorem{lemma}{Lemma}[section]

\DeclareMathOperator{\Tr}{Tr}

\numberwithin{equation}{section}

\begin{document}
\thispagestyle{empty}
\begin{center} 
{\Large   Imperfect Bose Gas\\[6pt]
 with Attractive Boundary Conditions
  }
 \\[30pt]
 
{\large
    L.~Vandevenne
       \footnote{Email: {\tt lieselot.vandevenne@fys.kuleuven.ac.be}},
    and A.~Verbeure
       \footnote{Email: {\tt andre.verbeure@fys.kuleuven.ac.be}}}
\\[30pt]
Instituut voor Theoretische Fysica,
Katholieke Universiteit Leuven,
Celestijnenlaan~200D,
B-3001 Leuven, Belgium\\[25pt]
{February, 2005}\\[25pt]
\end{center}
\begin{abstract}\noindent
The free Bose gas with attractive boundary conditions is an interesting toy model for the study of Bose-Einstein Condensation (BEC), because one has BEC already in one dimension. Here we study for the first time the imperfect Bose gas with these boundary conditions and prove rigorously the occurence of condensation.
\\[15pt]
\end{abstract}


\section{Introduction}

For the free Bose gas with Dirichlet and Neumann boundary conditions, Bose-Einstein Condensation (BEC) is rigorously treated in \cite{LP,LW}. For the mean-field Bose gas with periodic boundary conditions BEC is rigorously proved and a detailed analysis of the thermodynamic limit is given in \cite{FV}. The proof is based on bounds on the correlation functions for equilibrium states, given in terms of the correlation inequalities \cite{FV2,FV3}. The subtle point in this proof is the analysis of the singularity around the zero-mode.\\
If one considers attractive boundary conditions instead of periodic boundary conditions, the problem changes drastically. The free Bose gas with attractive boundary conditions is extensively studied in \cite{R,LW}. Due to the gap in the one-dimensional one-body problem, one has Bose-Einstein Condensation in all dimensions $\nu\geq 1$. An important result is the fact that the condensation is a surface effect. In \cite{VVZ} it is computed that the condensate is localized at a logarithmic distance from the boundary.\\
The subject of this note is to proceed with the imperfect Bose gas with attractive boundary conditions, \ie the free Bose gas with attractive boundary conditions plus a mean field term. The first problem that occurs is to express this mean field term in momentum space diagonalizing the kinetic energy (the free Bose gas part). As the spectrum of the latter one has two strictly negative eigenvalues, say $\epsilon_L(0)$ and $\epsilon_L(1)$, separated from the rest of the spectrum, then one can discuss the corresponding number operators $N_0$ and $N_1$ as being added to the total number operator in the interaction or not. In section \ref{interact_term} we argue why they should not be present. The argument is essentially based on the fact that we want a space homogeneous mean field term. The model is defined in \ref{hamiltonian}.\\
In section \ref{condensatie} we give a completely rigorous proof of the occurence of Bose-Einstein Condensation for the imperfect Bose gas with attractive boundary conditions. We perform all details only in dimension $\nu =1$. From dimension $\nu\geq 2$ on, the proof becomes technically more tedious. In particular, because of the fact that the condensate is located near the boundaries, for higher dimensions the thermodynamic limit for hypercubic boxes is not very suitable nor realistic and should in stead be taken with increasing absorbing balls. But we leave this extra exercise for a later occasion in which we consider the problem of the shape-dependence.\\
In the one-dimensional case we remark that the condensation is equally distributed over the two negative energy levels. The condensation is localized in the same area as for the free Bose gas with attractive boundary conditions, see \cite{R}.


\section{The Model}

\subsection{Attractive Boundary conditions}

If one considers a free gas of bosons in an interval $\left[-L/2,L/2\right]$ of length $L$, then the energy levels are determined by the one-dimensional Schr\"odinger equation (with units $\frac{\hbar^2}{2m}=1$) 
\begin{equation*}
-\Delta \phi = \epsilon_L \phi^L ,
\end{equation*}
with boundary
conditions:
$$\left\{
\begin{array}{lll}
\left(\displaystyle\frac{d\phi}{dx}-\sigma\phi\right)_{x=-L/2} & = & 0 ,\\
\left(\displaystyle\frac{d\phi}{dx}+\sigma\phi\right)_{x=L/2} & =
& 0 ,
\end{array} \right.$$
where $\sigma<0$.\\
If one considers these attractive boundary conditions, the spectrum consists of two negative eigenvalues tending to the same limit $-\sigma^2$ (when $L\rightarrow \infty$) and an infinite number of positive
eigenvalues (for $L|\sigma|>2$): $\epsilon_L(k)$ for $k=0,1,2,\ldots$, where
$$\epsilon_L(0) < \epsilon_L(1) <
0 < \epsilon_L(2) < \epsilon_L(3) < \ldots  ,$$
$$\epsilon_L(0)=-\sigma^2- O(\e^{-L|\sigma|}) ,$$
$$\epsilon_L(1)=-\sigma^2+O(\e^{-L|\sigma|}) ,$$
\begin{equation}\label{spect}
k\geq 2:\ \ \left(\frac{(k-1)\pi}{L}\right)^2 < \epsilon_L(k) <
\left(\frac{k\pi}{L}\right)^2 .
\end{equation}
The corresponding eigenfunctions $\{\phi_k^L\}_{k \in\N}$ are given by
\begin{eqnarray}
\phi_0^L(x) & = &
\sqrt{\frac{2}{L}}\left(1+\frac{\sinh(L|\sigma|)}{L|\sigma|}\right)^{-1/2}
\cosh(-|\sigma|x) ,\nonumber\\ \phi_1^L(x) & = &
\sqrt{\frac{2}{L}}\left(-1+\frac{\sinh(L|\sigma|)}{L|\sigma|}\right)^{-1/2}
\sinh(-|\sigma|x) ,\nonumber\\ \phi_k^L(x)& = & \left\{
\begin{array}{ll}
\sqrt{\displaystyle\frac{2}{L}}\left(1+\displaystyle\frac{\sin
(\sqrt{\epsilon_L(k)}L)}
{\sqrt{\epsilon_L(k)}L}\right)^{-1/2}\cos(\sqrt{\epsilon_L(k)}x) ,
\qquad & \mbox{for $k$ \ even} ,\\
\sqrt{\displaystyle\frac{2}{L}}\left(1-\displaystyle\frac{\sin
(\sqrt{\epsilon_L(k)}L)}
{\sqrt{\epsilon_L(k)}L}\right)^{-1/2}\sin(\sqrt{\epsilon_L(k)}x) ,
\qquad & \mbox{for $k$ odd} .
\end{array}\right.\nonumber
\end{eqnarray}

\subsection{Hamiltonian}\label{hamiltonian}

We consider a one-dimensional system of identical bosons on an interval $[-\frac{L}{2},\frac{L}{2}]\subset\R$ with attractive boundary conditions.
The model is specified by the local Hamiltonians $H_{L, MF}^{\s}$ on the boson Fock space $\mathcal{F}_{L, B}$:
\begin{equation}\label{ham}
H_{L, MF}^{\s} = T_{L}^{\s} + \frac{\l}{2}\frac{\tilde{N}_L^2}{L}
\end{equation}
where $T_{L}^{\s}$ is the kinetic energy operator with the $\epsilon_L(k)$ the eigenvalues (\ref{spect}) of the free Laplacian with attractive boundary conditions: 
\begin{equation*}
T_{L}^{\s} = \sum_{k\in\N}\epsilon_L(k) a_{k}^\ast a_{k}
\end{equation*}
The operators $a_k^\ast=a^\ast(\phi_L^k)$ and $a_k=a(\phi_L^k)$ are the Bose creation and annihilation operators with the testfunctions $\phi_L^k$ the above eigenfunctions of the free Laplacian with attractive boundary conditions. The total particle number operator is denoted by $N_L=\sum_{k\in\N}N_k = \sum_{k\in\N} a_{k}^\ast a_{k}$, and the particle number operator corresponding to the positive spectrum by $\tilde{N}_L$:
\begin{equation*}
\tilde{N}_{L} = \sum_{k=2}^{\infty}a_{k}^\ast a_{k}
\end{equation*}
We consider a positive coupling constant $\l \in \R^+$ for the sake of thermodynamic stability, see \cite{FV}.

\subsection{The Interaction Term}\label{interact_term}

Remark that the interaction in the Hamiltonian $H_{L, MF}^{\s}$ is not of the usual form $\frac{\l}{2}\frac{N_L^2}{L}$, with
\begin{eqnarray}
N_L & = & \int_{-L/2}^{L/2} \, dx \, a^\ast(x)a(x)\label{number}\\
 & = & \sum_{k\in\N}a_k^\ast a_k,\nonumber 
\end{eqnarray}
but of the form $\frac{\l}{2}\frac{\tilde{N}_L^2}{L}$, where
\begin{equation*}
\tilde{N}_L = \sum_{k=2}^\infty a_k^\ast a_k.
\end{equation*}
The reason for choosing $\tilde{N}_L$ in stead of $N_L$ is the breaking of the spatial translation invariance in the two terms with $k=0$ and $k=1$. Moreover the latter two terms yield also gauge symmetry breaking under the effect of space translations. Using the straightforward computation
\begin{equation*}
\cosh(|\s|(x+a)) = \cosh(|\s|x)\e^{-|\s|a}+\e^{|\s|x}\sinh(|\s|a)
\end{equation*}
for $a\in\R$, we get (for large $L$):
\begin{eqnarray*}
\tau_a (a_0^\ast) & \approx & a_0^\ast \e^{-|\s|a}+(a_0^\ast - a_1)\sinh(|\s|a)\\
\tau_a (a_0) & \approx & a_0 \e^{-|\s|a}+(a_0 - a_1^\ast)\sinh(|\s|a)
\end{eqnarray*}
where $\tau_a$ is the translation automorphism over the distance $a\in\R$.
One gets a similar expression for $\tau_a (a_1^\sharp)$.\\
In particular, the terms $\tau_a (N_0)$ and $\tau_a (N_1)$ are not translation invariant and diverge exponentially for $|a|\rightarrow\infty$ as
\begin{equation*}
\tau_a (a_0^\ast a_0) \approx \e^{2|\s||a|}a_0^\ast a_0 
\end{equation*}
moreover, these terms break the gauge symmetry.\\
Remark also that on the other hand the particle number operator $\tilde{N}_L$ is a good local approximation of $N_L$ (\ref{number}) for all translation invariant states, such that the interaction term $\frac{\l}{2}\frac{\tilde{N}_L^2}{L}$ is the appropriate mean field term. 


\section{Bounds on the correlation function and condensation}\label{condensatie}

Our aim is to find the equilibrium states of the system in the grand canonical ensemble. The equilibrium state $\omega_L$ at inverse temperature $\b$ is characterized by the following correlation inequality for all $L$, see \cite{FV2,FV3}
\begin{equation}\label{corr_ineq}
\b \omega_L(X^\ast [H_{L, MF}^{\s}-\mu_L N_L,X]) \geq \omega_L(X^\ast X)\ln \frac{\omega_L(X^\ast X)}{\omega_L(X X^\ast)}
\end{equation}
for all local observables $X$, where $\omega_{L}(\cdot)$ is the grand canonical equilibrium state at chemical potential $\mu_L$ and inverse temperature $\b$:
\begin{equation*}
\omega_{L}(X) = \frac{\Tr_{\mathcal{F}_{L, B}}X\exp\{-\b (H_{L, MF}^\s -\mu_L N_L)\}}{\Tr_{\mathcal{F}_{L, B}}\exp\{-\b H_{L, MF}^\s\}}
\end{equation*}
with $\mathcal{F}_{L, B}$ the boson Fock space over $\mathcal{L}^2([-L/2,L/2])$.\\
Concerning the thermodynamic limit ($L\rightarrow\infty$), we perform this limit keeping the total density $\rho$ constant. Therefore the chemical potential $\mu_L$ is now determined by the particle density $\rho$ and is the solution of the particle density equation: for each given density $\rho$ we have
\begin{equation}\label{part_density}
\rho = \frac{\omega_L(N_L)}{L}
\end{equation}
From the correlation inequality (\ref{corr_ineq}) follows immediately the inequality
\begin{equation}\label{corr_ineq2}
\omega_L\left(\big[X^\ast,[H_{L, MF}^\s -\mu_L N_L , X]\right]\big)\geq 0 .
\end{equation}
In this section we focus on the proof of the condensation for the model $H_{L, MF}^{\s}$ (\ref{ham}). In this proof we need bounds which we derive from the correlation inequality (\ref{corr_ineq}) for some specific observables $X$'s. Due to the special character of the spectrum (\ref{spect}), it is necessary to distinguish between products of creation and annihilation operators $a^\sharp_k$ in the $0$- or $1$-mode, and those in the $k$-mode (with $k\geq 2$).\\
The first Lemma is valid for all $k\in\N$.
\begin{lemma}\label{Nj_Nk}
If $j \neq k_i$ $(i = 1,\ldots ,m)$, $k_i \neq k_{i'}$ for $i \neq i'$ and 
$m,n_i \in \N_0$ for $i = 1,\ldots ,m$, then
\begin{eqnarray}
\lefteqn{\e^{\b (\epsilon_L(j)-\epsilon_L(k_1))}\omega_L\left(N_{j}
(N_{k_1}+1)^{n_1} (N_{k_2})^{n_2} \ldots (N_{k_m})^{n_m}\right)}\nonumber\\
 & = &  \omega_L\left((N_{j}+1)(N_{k_1})^{n_1} (N_{k_2})^{n_2} \ldots 
 (N_{k_m})^{n_m}\right)\label{lemma1}
\end{eqnarray}
\end{lemma}
\textit{Proof:}
The proof follows from the correlation inequality (\ref{corr_ineq}) by taking $X$
successively equal to
\begin{equation*}
a_{j}^\ast a_{k_1}\left((N_{k_1})^{n_1-1} (N_{k_2})^{n_2} \ldots 
 (N_{k_m})^{n_m}\right)^{1/2}
\end{equation*}
and
\begin{equation*}
a_{k_1}^\ast a_{j}\left((N_{k_1})^{n_1-1} (N_{k_2})^{n_2} \ldots 
 (N_{k_m})^{n_m}\right)^{1/2}
\end{equation*}
\hfill $\square$\\
For the chemical potential $\mu_L$, we find the same upperbound as in the case of the free Bose gas with attractive boundary conditions.
\begin{lemma}\label{mu}
For $k=0,1$, one has
\begin{equation*}
\mu_L \leq \epsilon_L(k)
\end{equation*}
and
\begin{equation}\label{mu_expr}
\mu = \lim_{L\rightarrow\infty}\mu_L \leq -\s^2
\end{equation}
\end{lemma}
\textit{Proof:}
Take $X = a_{k}^\ast$ where $k=0,1$ in the inequality (\ref{corr_ineq2}).\\ 
The second inequality (\ref{mu_expr}) is obtained by taking the thermodynamic limit $L$ tending to infinity.
\hfill $\square$

\begin{lemma}\label{Nk_B}
For each $k=0,1$ and $\mu_{L}<-\s^2$, we have:
\begin{equation}
\omega_{L}(N_{k}) = \frac{1}{\e^{\b(\epsilon_L(k)-\mu_L)}-1}
\end{equation}
\end{lemma}
\textit{Proof:}
The result follows from the inequality (\ref{corr_ineq}) by taking $X$ successively equal to $a_{k}$ and $a_{k}^\ast$ with $k=0,1$.
\hfill $\square$

\begin{lemma}\label{corr_ineq_Nk}
For each $n\in\N$ and $k\geq 2$, we have
\begin{equation*}
\b\omega_{L}\left(-\epsilon_{L}(k)N_{k}^{n+1}+\mu_{L} N_{k}^{n+1}-\l \frac{\tilde{N}_{L}}{L}N_{k}^{n+1} +\frac{\l}{2}\frac{N_{k}^{n+1}}{L}\right)\geq \omega_L(N_{k}^{n+1})\ln \frac{\omega_L(N_{k}^{n+1})}{\omega_L((N_{k}+1)^{n+1})}
\end{equation*}
\end{lemma}
\textit{Proof:} The result follows from the correlation inequality (\ref{corr_ineq}) with $X=a_k N_k^{n/2}$.
\hfill $\square$\\ \\
In order to prove condensation, we need a convenient upperbound for $\omega_{L}(N_{k})$ for all $k\in\N$. This bound is derived in the following Lemma.
\begin{lemma}\label{Nk}
For each $k\geq 2$ we have
\begin{equation*}
\omega_{L}(N_{k}) \leq \frac{1}{\e^{c_{k}(L)}-1}
\end{equation*}
where
\begin{equation}\label{ck}
c_{k}(L) = \b\left(\epsilon_{L}(k)+\s^2-\frac{\l}{2L}-o(\e^{-L|\s|})\right)
\end{equation}
\end{lemma}
\textit{Proof:}
By Lemma \ref{corr_ineq_Nk} with $n=0$:
\begin{equation*}
\omega_{L}(N_{k})\ln\frac{\omega_{L}(N_{k})}{\omega_{L}(N_{k})+1} \leq \b\omega_L\left(-\epsilon_{L}(k)N_{k} + \mu_{L} N_{k} - \l\frac{\tilde{N}_{L}}{L}N_{k} + \frac{\l}{2}\frac{N_{k}}{L}\right)
\end{equation*}
From Lemma \ref{mu} we know that $\mu_L \leq -\s^2 - o(\e^{-L|\s|})$. It is also easy to see that $\omega_L(\tilde{N}_L N_k) \geq 0$. This leads to
\begin{equation*}
\omega_{L}(N_{k})\ln\frac{\omega_{L}(N_{k})}{\omega_{L}(N_{k})+1} \leq -\b\left(\epsilon_{L}(k) + \s^2 + o(\e^{-L|\s|}) + \frac{\l}{2L}\right)\omega_\L(N_k)
\end{equation*}
which gives us immediately the result.
\hfill $\square$\\ \\
Using the results of the previous Lemma's, we are now ready to prove the existence of condensation in the two lowest energy levels.
\begin{theorem}\label{condensation}
Let $\rho_{cond}$ be equal to 
\begin{equation*}
\rho_{cond} = \lim_{L\rightarrow\infty} \frac{1}{L}\omega_L\left(N_0 + N_1\right).
\end{equation*} 
Then
\begin{equation}\label{condensation_expr}
\rho_{cond} \geq \rho - \frac{1}{\pi}\int_{0}^{\infty}\, dk \frac{1}{\e^{\b(k^2+\s^2)}-1}
\end{equation}
where $\rho_{cond}$ is the density of the condensate. The condensate density is localized in the $2$ lowest energy levels.
\end{theorem}
\textit{Proof:}
From the definition of the particle density $\rho$ (\ref{part_density}), we have
\begin{equation*}
\frac{1}{L}\omega_L\left(N_0 + N_1\right) = \rho - \frac{1}{L} \sum_{k=2}^\infty\omega_{L}(N_k)
\end{equation*}
By using the estimate of Lemma \ref{Nk}, one gets
\begin{equation*}
\frac{1}{L}\omega_L\left(N_0 + N_1\right) \geq \rho - \frac{1}{L} \sum_{k=2}^\infty \frac{1}{\e^{c_{k}(L)}-1}
\end{equation*}
with the $c_k(L)$'s as in (\ref{ck}).\\
Taking the thermodynamic limit $L\rightarrow\infty$ gives us the result (\ref{condensation_expr}).
\hfill $\square$\\ \\
Clearly (\ref{condensation_expr}) shows condensation. Indeed, remark that the integral is convergent for all $\s\neq 0$ and that it decreases for $\b$ increasing. Hence for $\rho$ large enough or for $\b$ large enough, it follows that the condensate density $\rho_{cond}$ is strictly positive.\\
Finally we derive a result about the type of condensation. We prove that the condensate density is realized in both the two lowest energy modes, with equal weight in the thermodynamic limit.

\begin{theorem}
\begin{itemize}
\item[(i)]The condensate is equally distributed on the two lowest energy levels. 
\item[(ii)]From this, one can compute the asymptotics of the chemical potential $\mu_L$ for large $L$:
\begin{equation}\label{mu_asymp}
\mu_L = -\s^2 - \frac{2}{\b\rho_{cond} L} + o(L^{-1})
\end{equation}
\end{itemize}
\end{theorem}
\textit{Proof:}
\begin{itemize}
\item[(i)] From Lemma \ref{Nj_Nk}, with $m=1$, $j=1$ and $k=0$, we get
\begin{equation*}
\left(\e^{\b(\epsilon_L(1)-\epsilon_L(0))}-1\right)\omega_L(N_1 N_0) + \e^{\b(\epsilon_L(1)-\epsilon_L(0))}\omega_L(N_1) = \omega_L(N_0)
\end{equation*}
Using the spectral properties
\begin{eqnarray*}
\epsilon_L(0) & = & -\sigma^2- O(\e^{-L|\sigma|})\\
\epsilon_L(1) & = & -\sigma^2+O(\e^{-L|\sigma|})
\end{eqnarray*}
such that
\begin{equation*}
\epsilon_L(1)-\epsilon_L(0)\approx o(\e^{-L|\s|}),
\end{equation*}
then
\begin{equation*}
\frac{\omega_\L(N_0)}{L} = \frac{\omega_\L(N_1)}{L} + o(\e^{-L|\s|}).
\end{equation*}
Taking the thermodynamic limit $L\rightarrow\infty$, one gets (i).
\item[(ii)] From Lemma \ref{Nk_B} and since we know that the condensate is equally distributed on the two lowest energy levels, we have
\begin{equation*}
\lim_{L\rightarrow\infty}\frac{\omega_L(N_{0})}{L} = \lim_{L\rightarrow\infty} \frac{1}{L}\frac{1}{\e^{\b(\epsilon_L(0)-\mu_L)}-1} = \frac{\rho_{cond}}{2} 
\end{equation*}
Series expansion of this expression with respect to the quantities $\epsilon_L(0)-\mu_L$ and $\epsilon_L(0)=-\sigma^2- O(\e^{-L|\sigma|})$, gives us the asymptotics of $\mu_L$.
\end{itemize}
\hfill $\square$



\begin{thebibliography}{}

\bibitem{FV2}M.~Fannes and A.~Verbeure,
{\em Correlation Inequalities and Equilibrium States II}~; Commun. Math. Phys. \textbf{57}, 165-171 (1977)

\bibitem{FV3}M.~Fannes and A.~Verbeure,
{\em Global Thermodynamic Stability and Correlation Inequalities}~; J. Math. Phys. \textbf{19}, 558-560 (1978)

\bibitem{FV}M.~Fannes and A.~Verbeure,
{\em The Condensed phase of the Imperfect Bose Gas}~; J. Math. Phys. \textbf{21}(7), 1809-1818 (1980)

\bibitem{LW}J.~Landau and I.F.~Wilde,
{\em On the Bose-Einstein Condensation of an Ideal Gas}~; Commun. Math. Phys. \textbf{70}, 43-51 (1979)

\bibitem{LP}J.T.~Lewis and J.V.~Pul\'e,
{\em The Equilibrium State of the Free Boson Gas}~; Commun. Math. Phys. \textbf{36}, 1-18 (1974)

\bibitem{R}D.W.~Robinson,
{\em Bose-Einstein Condensation with Attractive Boundary
Conditions}~; Commun. Math. Phys. \textbf{50}, 53-59 (1976)

\bibitem{VVZ}L.~Vandevenne, A.~Verbeure and V.A.~Zagrebnov,
{\em Equilibrium States for the Bose Gas}~; J. Math. Phys. \textbf{45}(4), 1606-1622 (2004)

\end{thebibliography}
\end{document}